\documentclass[aps,prl,twocolumn,superscriptaddress]{revtex4-2}

\pdfoutput=1

\usepackage{graphics}
\usepackage{amssymb}
\usepackage{amsmath}
\usepackage{hyperref}
\usepackage{bbold}
\usepackage{tikz}
\usepackage{ulem}
\usepackage{booktabs, siunitx}
\usepackage{float}

\begin{document}
\title{Exploring Quantum Spacetime with Topological Data Analysis}

\author{J. van der Duin}
\email{jesse.vanderduin@ru.nl}
\affiliation{Institute for Mathematics, Astrophysics and Particle Physics, Radboud University, Heyendaalseweg 135, 6525 AJ Nijmegen, The Netherlands.}

\author{R. Loll}
\email{r.loll@science.ru.nl}
\affiliation{Institute for Mathematics, Astrophysics and Particle Physics, Radboud University, Heyendaalseweg 135, 6525 AJ Nijmegen, The Netherlands.}
\affiliation{Perimeter Institute, 31 Caroline Street North, Waterloo, ON, N2L 2Y5, Canada.}

\author{M. Schiffer}
\email{marc.schiffer@ru.nl}
\affiliation{Institute for Mathematics, Astrophysics and Particle Physics, Radboud University, Heyendaalseweg 135, 6525 AJ Nijmegen, The Netherlands.}

\author{A. Silva}
\email{agustin.silva@ru.nl}
\affiliation{Institute for Mathematics, Astrophysics and Particle Physics, Radboud University, Heyendaalseweg 135, 6525 AJ Nijmegen, The Netherlands.}

\begin{abstract}
In a novel application of the tools of topological data analysis (TDA) to nonperturbative quantum gravity, we introduce a new class of observables that allows us to assess 
whether quantum spacetime really resembles a ``quantum foam" near the Planck scale. 
The key idea is to investigate the Betti numbers of coarse-grained path integral histories, regularized in terms of dynamical triangulations,  
as a function of the coarse-graining scale. In two dimensions our analysis exhibits the well-known fractal structure of Euclidean quantum gravity.
\end{abstract}

\maketitle

\section{Quantum gravity, from the nonperturbative end}
\label{sec:intro}

The better we understand how to computationally access the nonperturbative realm of quantum gravity with quantum field theoretic tools, using dynamical lattices \cite{ency} or 
the functional renormalization group \cite{Reuter2019}, the more we can focus on interesting physical and conceptual questions. 
A key challenge is to understand the physical nature of the strongly quantum-fluctuating \textit{quantum spacetime} near the Planck scale, and to find 
the observables most suited to probing it. This is a road less travelled, since the required computational techniques beyond perturbation theory are not (yet) part of many
practitioners' toolbox. Nevertheless, their use is already producing new and promising \textit{quantitative} results in a hitherto inaccessible Planckian regime, and may provide
a unique gateway to understanding what quantum gravity is fundamentally about. 

The go-to method for analyzing a quantum field theory in its nonperturbative regime is by putting it on a spacetime lattice, as exemplified by the formidable successes of 
lattice QCD \cite{Rothe2012}. However, the rigid character of the lattices used there clashes directly with the dynamical, curved nature of spacetime in gravity,
a key problem of \textit{lattice quantum gravity} already articulated long ago \cite{Smolin1978}. Several conceptual and technical breakthroughs were needed to
arrive at a viable lattice theory to compute the nonperturbative gravitational path integral 
\begin{equation}
Z=\int {\cal D}[g]\ \mathrm{e}^{\, i S^{\mathrm{EH}}[g]}
\label{pathint}
\end{equation}
over diffeomorphism equivalence classes $[g_{\mu\nu}]$ of spacetime metrics $g_{\mu\nu}$, where
\begin{equation}
 \;\;\;\; S^{\mathrm{EH}}[g]= \frac{1}{16\pi G_{\rm N}}\, \int_M d^4 x\, \sqrt{|\det(g)|}\, (R -2 \Lambda )
\label{seh}
\end{equation}
is the Einstein-Hilbert action with a cosmological term. The resulting ``lattice quantum gravity 2.0" \cite{Loll2025}, based on causal dynamical triangulations (CDT), is now available, has 
been thoroughly tested and has delivered many new results (see \cite{review1,review2} for reviews). 
It combines
(i) \textit{dynamical} instead of fixed hypercubic lattices, reflecting the dynamical nature of spacetime geometry, 
(ii) an \textit{exact} relabelling symmetry, the lattice analogue of diffeomorphism symmetry in the continuum, and 
(iii) a \textit{Wick rotation} for curved lattice spacetime configurations, which has no known counterpart in the continuum. The latter is essential for employing 
powerful Monte Carlo Markov Chain methods to evaluate the lattice-regularized version of the \textit{Lorentzian} path integral (\ref{pathint}), after using this analytic 
continuation.\footnote{Note that this set-up is inequivalent to the conceptually and computationally rather ill-defined Euclidean gravitational path integral in four dimensions, 
which is at the core of so-called Euclidean quantum gravity, see e.g.\ \cite{Ambjorn2024}.}

\section{Introducing effective homology}
\label{sec:effhom}

Modern lattice quantum gravity will also be the setting for this letter. We will introduce a new class of observables to characterize the local structure of quantum spacetime, which
is generated dynamically by the nonperturbative gravitational path integral (\ref{pathint}). Our construction is inspired by concepts and techniques from the field of 
topological data analysis (TDA), whose main application is the characterization of very large data sets in terms of certain geometric and topological properties one can associate with them (see \cite{Carlsson2009,Munch2017} for an introduction). More specifically, we will tap into ideas from persistent homology, where one studies the homology of a 
simplicial complex associated with a data set, as a function of some scale parameter \cite{roadmap}. It will allow us to use GUDHI \cite{gudhi}, one of a number of open-source TDA libraries, as a powerful technical tool to compute the homological properties of large triangulations.

We propose to investigate the \textit{effective homology} of quantum spacetime, which we define as the homology of a coarse-grained version of the quantum geometry, 
as a function of the coarse-graining scale. More specifically, the quantum observables we will measure are the Betti numbers of the coarse-grained 
quantum geometry. Recall that the Betti numbers capture topological features like connected components, loops and voids.
Concretely, we propose to coarse-grain the triangulated path-integral configurations in the lattice implementation of (\ref{pathint}) 
by an integer scale $\delta=2,3,4,\dots$, which sets the length of an edge in the coarse-grained lattice, in units of the original lattice edges.

Note that the configurations summed over in the path integral all have the same, fixed topology (in four dimensions usually that of $S^1\times S^3$, 
with the time direction cyclically identified), but 
that a coarse-graining will in general change this topology, in a way that depends on $\delta$ and is characteristic of the underlying quantum geometry. 
It will enable us to explore to what extent this generalized geometry resembles a quantum spacetime foam \cite{Carlip2022}, 
a frequently invoked image for ``whatever becomes
of spacetime at the Planck scale". Its local structure is often conjectured to be highly nontrivial topologically, like a bath of foam bubbles or riddled with wormholes. 
In the lattice regularization, such a structure by construction is not present
at the scale of the lattice cut-off (for $\delta=1$), where the topology is fixed by fiat\footnote{Note that lattice properties at or very near to the cut-off are usually discarded, 
because they reflect lattice artefacts rather than properties of any continuum limit.}, but can in principle ``emerge" when considering larger resolutions $\delta$.  
Beyond characterizing quantum geometry as such, we also expect that the effective topology at coarse-graining scale $\delta$ reflects properties that will be felt by a matter probe of linear extension $\delta$ or a wavelike excitation of wave length $\delta$.

In what follows, we will demonstrate the viability of our new methodology in the well-known toy model of two-dimensional Euclidean quantum gravity, 
also known as Liouville gravity \cite{David1984,Kazakov1985} (see \cite{Budd2022} for a recent review). 
We describe the steps for the coarse-graining procedure on the two-dimensional triangulations of the regularized path integral and
the numerical measurement results obtained for the expectation values $\langle \beta_i(\delta)\rangle$, $i=0,1,2$, of the Betti numbers. We then relate the nontrivial behaviour of
$\beta_2$ to the fractal nature of the quantum geometry, as quantified by the string susceptibility $\gamma_\mathrm{str}$, a universal scaling exponent governing the distribution
of baby universes \cite{Jain1992}. A companion work \cite{followup} contains a complete set of technical details of the construction in two dimensions, and an application where we use the effective homology to assess the homogeneity of both Euclidean and Lorentzian quantum geometry.

\section{Path integral set-up}
\label{sec:setup}

Working in two Euclidean dimensions and using a lattice regularization in terms of dynamical triangulations, the path integral (\ref{pathint}), (\ref{seh}) assumes the concrete form 
\begin{equation}
Z^\mathrm{eu}(\lambda)=\sum_{T} \tfrac{1}{C(T)}\, \mathrm{e}^{-\lambda N_2(T)} ,
\label{pi2dedt} 
\end{equation}
where the sum is over all equilateral triangulations $T$ with the topology of a two-sphere, $\lambda$ denotes the (bare) cosmological coupling constant,
the discrete volume $N_2(T)$ counts the number of triangles in $T$, and $C(T)$ is the order of the automorphism group of $T$, counting the number of ways in which $T$
can be mapped onto itself while preserving its neighbourhood relations. Without loss of generality, the ensemble $\{T\}$ used here is that of triangulations dual to trivalent graphs with self-energies, but without tadpoles \cite{followup,Loll2024}. The curvature term has been dropped from the action since it is a topological invariant independent of $T$. 

\begin{figure}[t!]
\centering
\includegraphics[width=0.49\textwidth]{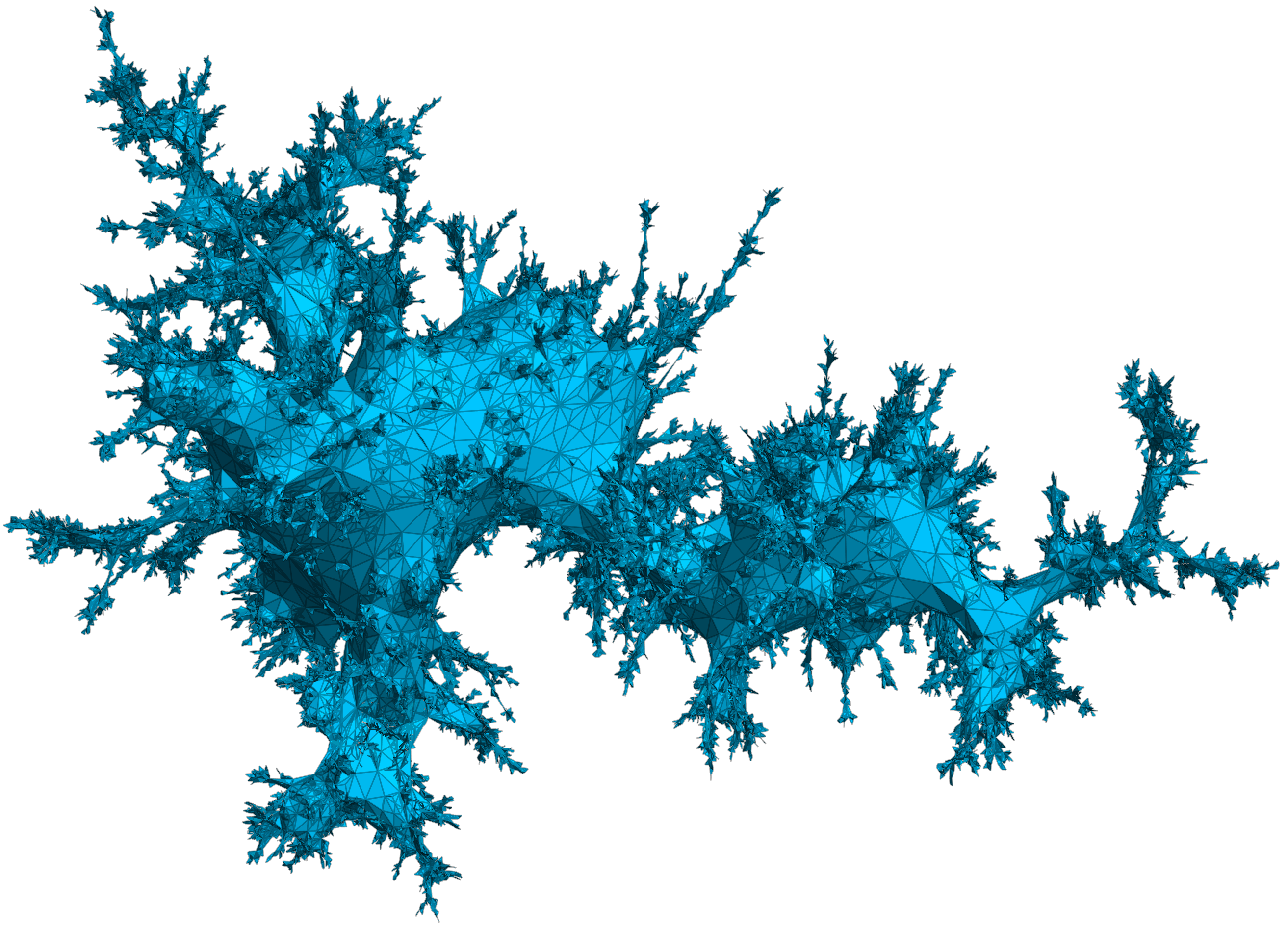}\\
\caption{Typical configuration contributing to the Euclidean path integral (\ref{pi2dedt}) over 2D dynamical triangulations, for $N_2=100k$. 
}
\label{fig:50k-base}
\end{figure}

Note that each triangulation $T$ is a piecewise flat, metric space, whose local, intrinsic curvature properties are determined by its neighbourhood relations, i.e.\ by how its flat
equilateral triangles are glued together pairwise along shared edges. 
These spaces also carry a natural notion of geodesic (link) distance, defined between pairs $(v_1,v_2)$ of their vertices as the number of links (= edges)
in the shortest path connecting $v_1$ and $v_2$. A typical member of the ensemble of dynamically triangulated spheres is depicted in Fig.\ \ref{fig:50k-base}, illustrating the
well-known fact that the geometry does not at all resemble that of a round continuum sphere or any other ``nice" classical space. 

The Euclidean path integral (or partition function) (\ref{pi2dedt}) can be computed analytically, involving a fine-tuning of $\lambda$ to its critical value, to yield a continuum theory of 
2D Euclidean quantum gravity in the universality class of Liouville quantum gravity. However, since the Betti number observables $\beta_i(\delta)$ depend on the underlying 
configurations $T$ in a complicated way, we will compute them numerically, using a direct Monte Carlo sampling \cite{followup}. As usual \cite{Ambjorn1997}, we will evaluate all
expectation values 
\begin{equation}
\langle {\cal O}\rangle_{N_2} =\frac{1}{\tilde{Z}^\mathrm{eu}(N_2)}   \sum_{T|_{N_2}} \tfrac{1}{C(T)}\, {\cal O}(T)
\label{expfix}
\end{equation}
of observables ${\cal O}(T)$ in ensembles of constant volume $N_2$ and in the limit as $N_2$ becomes large. The corresponding partition function $\tilde{Z}^\mathrm{eu}(N_2)$ 
is related to (\ref{pi2dedt}) by
\begin{equation}
Z^\mathrm{eu}(\lambda)=\sum_{N_2} \mathrm{e}^{-\lambda N_2} \tilde{Z}^\mathrm{eu}(N_2),\;\;\;\;\; \tilde{Z}^\mathrm{eu}(N_2)=  \sum_{T|_{N_2}} \tfrac{1}{C(T)}.
\label{zvol}
\end{equation}
To estimate the expectation values $\langle \beta_i(\delta)\rangle_{N_2}$, we have measured $\beta_i(\delta)$, for $\delta\!\in\! [2,53]$, 
on several hundred thousand independent configurations $T$ for
each of eight volumes in the range $N_2\in[50k,400k]$. For a given triangulation $T$ and resolution $\delta$, the measurement involves a coarse-graining
procedure that produces an equilateral triangulation $T_\delta$ with triangles of edge length $\delta$, which then serves as an input for GUDHI.\footnote{Note that for $\delta=1$, 
the coarse-graining algorithm described in the next section reproduces the original triangulation, i.e.\ $T_1=T$.}  
The latter can proceed in a highly efficient way, because the computation of the Betti numbers of a simplicial complex over a finite field
amounts to a problem in linear algebra \cite{roadmap}.

\section{Coarse-graining of geometry}
\label{sec:coarse}

The coarse-graining of a configuration $T$ to arrive at a triangulation $T_\delta$, with correspondingly fewer triangles, proceeds in several steps:
\begin{itemize}
\item[(i)] select an evenly distributed subset ${\cal S}_\delta$ of all vertices of $T$, such that the link distance between nearest neighbours is of the order $\delta$, 
\item[(ii)] use ${\cal S}_\delta$ to construct a Voronoi decomposition of $T$, and 
\item[(iii)] construct the dual of the Voronoi decomposition, which is the searched-for coarse-grained (generalized) Delaunay triangulation $T_\delta$.
\end{itemize}
Since an understanding of the technical details of this procedure is not essential here, we will confine ourselves to a summary of the main points and
refer the interested reader to \cite{followup} for further details. 

To implement step (i) above, we use a construction loosely analogous to Poisson disk sampling.
Starting at a random vertex $v_0$, we determine the vertices lying inside a ball of radius $\delta$ and an annulus between the radii $\delta$ and $2\delta$ around $v_0$.
By suitably selecting vertices for ${\cal S}_\delta$ 
from these subsets and reiterating the process by constructing their $\delta$-balls and $\delta$-annuli, until the entire triangulation $T$ has been covered,
we end up with a subset ${\cal S}_\delta$ with the desired even spread.

\begin{figure}[t]
\centering
\includegraphics[width=0.45\textwidth]{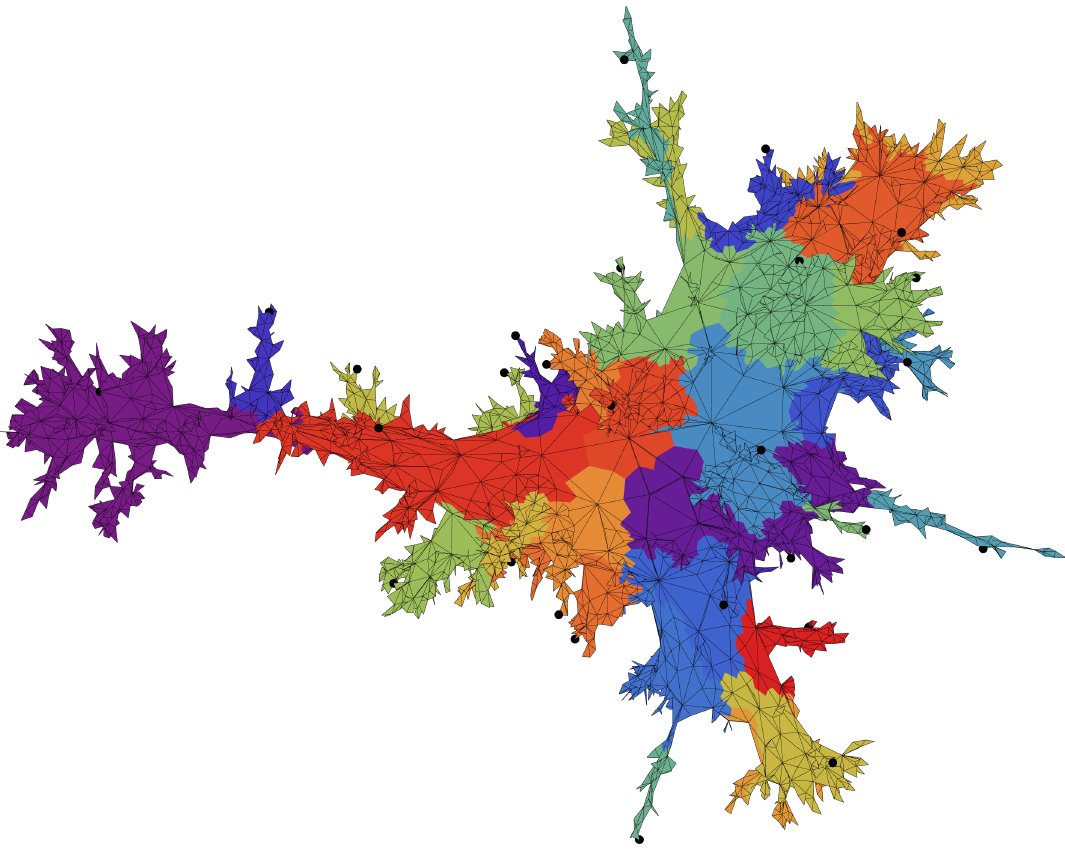}
\caption{Voronoi decomposition with resolution $\delta=8$ of a typical configuration, with $N_2=10k$. 
Thin black lines are those of the original triangulation $T$.
}
\label{fig:spheredelta8}
\end{figure}
Regarding step (ii), the Voronoi cell associated with a given vertex $v\in{\cal S}_\delta$ contains all vertices closer (in link distance) 
to $v$ than to any other $v'\in {\cal S}_\delta$. The cells are constructed by running a simultaneous breadth-first search, with the vertices of ${\cal S}_\delta$ as seeds.
After associating vertices to cells, there is a deterministic way -- involving a three-way colouring of the triangles of $T$ -- to associate all points of $T$ with the interior of a
cell, a boundary between two cells, or a triple point where three cells meet, see Fig.\ \ref{fig:spheredelta8}.
A main point to note here is that because of the irregularity of the underlying triangulation $T$, the possible shapes of the resulting individual Voronoi cells and the manner in which neighbouring
cells meet each other are generalized, compared to an analogous construction in the flat plane. Topologically, the cells need not be simple discs, but can have additional holes cut out, e.g.\ like an annulus. In addition, the boundary between two neighbouring cells can consist of two or more disconnected segments. 

These features have direct consequences for step (iii), where it turns out that the Delaunay triangulation $T_\delta$ dual to the Voronoi decomposition of (ii) is no longer a topological manifold. The elements that make up a Delaunay triangulation are the vertices dual to the Voronoi cells, the edges dual to the boundary segments between Voronoi cells, and the triangles dual to the trivalent vertices of the Voronoi decomposition. Their connectivities are inherited from those of $T$, at least on linear scales 
$\gtrsim\delta$. Broadly speaking, their metric properties are also inherited from $T$ by assigning uniform length $\delta$ to all edges in $T_\delta$ and declaring all triangles in $T_\delta$ as flat.

To illustrate why $T_\delta$ is no longer a manifold (i.e.\ does not look two-dimensional in the neighbourhood of every point), note that 
the building blocks of $T_\delta$ are not only triangles, but also ``loose edges". These are edges that do not belong to the boundary of any triangle, 
and are attached to the rest of the triangulation at one or both of their endpoints. It is easy to see how they can come about during coarse-graining. 
Suppose that the original $T$, which is a manifold by assumption, contains a ``thin neck". By this, we mean a
closed loop of $\ell$ edges, for small $\ell =2,3,4,\dots$, such that cutting $T$ open along the loop results in two components, each of which contains more than a minimal number of triangles, for some suitable, $\ell$-dependent definition of ``minimal". 
Typical path integral configurations of Euclidean 2D quantum gravity are full of thin necks of various sizes, as
illustrated by Figs.\ \ref{fig:50k-base} and \ref{fig:spheredelta8}. 

For a coarse-graining with resolution $\delta$, thin necks of length $\ell\lesssim\delta$ will typically lead to a 
``pinching" of the Delaunay triangulation, where the entire loop is effectively shrunk to a single vertex, which then becomes a non-manifold point of $T_\delta$. If the original
triangulation $T$ had a ``long thin neck", one that continued to be thin in the direction perpendicular to the loop, it will appear as a loose edge or a sequence of loose edges in the coarse-grained $T_\delta$.\footnote{Note that we do not allow parts of the triangulation to completely detach during coarse-graining.} 

As we will see next, these pinchings influence the behaviour of
the Betti numbers in a nontrivial way.

\section{Betti numbers: results}
\label{sec:betti}

\begin{figure}[t]
\centering
\includegraphics[width=0.49\textwidth]{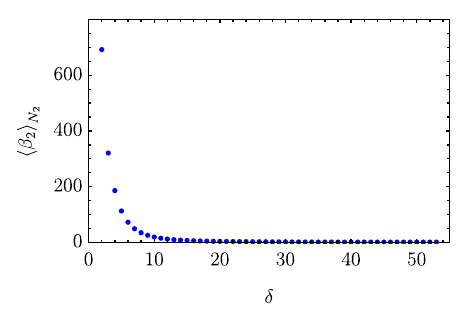}
\caption{Expectation value of the Betti number $\beta_2$,
as a function of the resolution $\delta\in [2,53]$ and for $N_2=200k$.}
\label{fig:betti}
\end{figure}

By construction, the Betti numbers of any triangulation $T$ are those of a two-sphere, namely $\beta_0\! =\! 1$ (a single connected component),
$\beta_1\! =\! 0$ (no non-contractible loops) and $\beta_2\! =\! 1$ (the sphere encloses a single two-dimensional ``hole"). Our Monte Carlo measurements on triangulations with volume 
$N_2\!\in\! [50k,400k]$ reproduce these numbers not just for the expectation values $\langle \beta_0(\delta)\rangle $ and $\langle\beta_1(\delta)\rangle$, but for each individual $T$ and in the studied
range $\delta\in[2, 53]$. This result is explained by the fact that the coarse-graining does not alter the connectivity and also cannot generate non-contractible loops. 

However, the measurements for $\langle \beta_2(\delta)\rangle$ deviate strongly from its classical value, as illustrated by the data shown in Fig.\ \ref{fig:betti}. 
For the smallest nontrivial coarse-graining $\delta\!=\!2$, 
$\langle \beta_2\rangle$ is very large ($692.221 \pm 0.040$ for the plot shown), decreases steeply for larger $\delta$, and gradually asymptotes to a value compatible with 1 at the end of the range considered.\footnote{The coarse-graining is stopped before all triangles disappear, which guarantees that $\beta_2\geq 1$.}  
What we see here is the
effect of the pinching process described above, which generates many more holes or ``bubbles" besides that of the original two-sphere.\footnote{A
\textit{bubble} denotes a part of the triangulation that is topologically a two-sphere and connected to the rest of the triangulation at a (pinching) vertex or along loose edges.}  
Their number is maximal for $\delta\! =\! 2$, since the volumes of these bubbles are relatively small and they rapidly disappear for larger resolutions, 
whenever $\delta$ becomes larger than their linear size. 

We conclude that we have found a nontrivial example of the type of effective homology we were looking for, where the behaviour of the Betti number $\beta_2$ after coarse-graining
captures a local feature of the underlying quantum geometry. Since the effect we found is associated with relatively small values of $\delta$, one could wonder whether this is
a short-distance lattice artefact irrelevant for the continuum theory. We will show next that this is not the case and instead our findings are related to 
well-known properties of 2D Euclidean quantum gravity.

\section{Recovering the string susceptibility} 
\label{sec:susc}

Our discussion of pinchings and thin necks is reminiscent of earlier work on 2D Euclidean quantum gravity involving so-called minimal-neck baby universes or ``minbus" \cite{Jain1992}.
Minimal necks are thin necks of length 3 in an ensemble of equilateral triangulations that obey slightly stricter regularity conditions\footnote{variably called 
combinatorial triangulations or simplicial manifolds, see e.g.\ \cite{Loll2024}; for the triangulations used by us the minimal neck length is 2}, but whose use in the path integral leads to the same continuum theory.  
Cutting open the triangulation along such a minimal neck will generically lead to a ``mother universe", where most of the
volume resides, and a much smaller minbu\footnote{When the baby universe consists of a single triangle only, it is not counted as a minbu.}. 
It has been shown that the distribution of minbu sizes $n$ for triangulations of fixed volume $N_2$ has a simple dependence on 
the so-called string susceptibility $\gamma_\mathrm{str}$ \cite{Jain1992}, and that measuring this distribution 
provides an efficient way to extract $\gamma_\mathrm{str}$ numerically \cite{Ambjorn1993}. 

This scaling exponent governs the subleading behaviour of the partition function $\tilde{Z}(N_2)$ for fixed volume, which for large $N_2$ is known analytically
to have the form
\begin{equation}
\tilde{Z}(N_2)\propto \mathrm{e}^{\lambda_0 N_2} N_2^{\gamma_\mathrm{str} -3} (1+{\cal O}(\tfrac{1}{N_2}))
\label{minbu1}
\end{equation}
for some (non-universal) constant $\lambda_0 >0$. 
From this one can derive that the expectation value $\langle b(n)\rangle_{N_2}$ of the number of 
minbus with volume $0\ll n < N_2/2$ 
\begin{equation}
\langle b(n)\rangle_{N_2}\propto n^{\gamma_\mathrm{str}-2} (N_2-n)^{\gamma_\mathrm{str}-2}.
\label{minbu2}
\end{equation}

The similarities with our construction raise the question of whether we can relate the behaviour of our bubbles in a quantitative way to the minbu analysis. 
This would require us to keep track of the volumes of the bubbles and consider their statistics. Despite the differences between the two set-ups, this turns
out to be possible and leads to a surprisingly close match when extracting the scaling parameter analogous to $\gamma_\mathrm{str}$. 

It is most natural to compare with the case where our necks have minimal length, i.e.\  $\delta\! =\! 2$, for which also the number of bubbles is maximal.\footnote{The 
mismatch between $\ell_\mathrm{min}\! =\! 2$ and $\ell_\mathrm{min}\! =\! 3$ is a lattice artefact, which should not play much of a role.}
The potentially most relevant difference comes from the fact that bubbles and minbus are different objects. The standard minbu analysis for a given triangulation
$T$ identifies all minimal necks and for each such neck records the volume of the smaller component (with $n\! <\! N_2/2$), regardless of whether it
contains other minimal necks and associated minbus \cite{Ambjorn1993}. By contrast, for a given coarse-grained triangulation $T_2$, bubbles are mutually exclusive, 
and adding up their volumes -- given in terms of coarse-grained triangle units -- one obtains the total volume of $T_2$.

The reason why one can nevertheless expect a similar volume distribution is the typical baby universe structure of the triangulations under consideration,
which we have investigated qualitatively \cite{followup}. It consists of a ``mother bubble" of large volume, which via pinching vertices and loose edges
is connected to a first generation of much smaller ``baby bubbles", which in turn can have further bubbles as offspring, and so on. However,
baby bubbles of second or higher generation are rare: less than 10\% of first-generation bubbles have further offspring, and even if they do, their volumes tend to
be small. It implies that many bubbles can be identified with minbus of the same or similar volume.    

\begin{figure}[t!]
\centering
\includegraphics[width=0.49\textwidth]{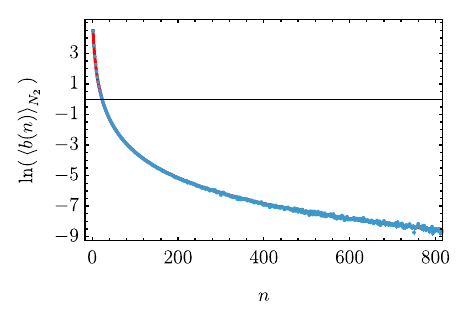}\\
\caption{Expectation value of the abundance of bubbles of volume $n$, for $\delta=2$ and $N_2=50k$, together with a best fit for the string susceptibility.}
\label{fig:50k_string_susc}
\end{figure}

We have measured the abundance $b(n)$ of bubbles of volume $n$ for several hundred thousand triangulations for each of the eight volumes $N_2\in [50,400k]$ considered and fitted the data to the functional form
\begin{equation}
\ln (\langle b(n)\rangle_{N_2}) = \alpha +(\gamma_\mathrm{str} -2) \ln \big(n(1-\frac{n}{N_2^{c}}) \big)+\frac{\kappa}{n},
\label{bufit}
\end{equation}
where $\alpha$ and $\kappa$ are fit parameters, $n$ and $N_2^c$ are measured in terms of coarse-grained volume units\footnote{We have verified that $N_2^c$ is roughly proportional 
to $N_2$, with a Gaussian spread.}, $\kappa/n$ is a finite-size correction term, and a bubble size of 5\% of the total volume $N_2^c$ was chosen as upper bound for the fitting window.
For illustration, Fig.\ \ref{fig:50k_string_susc} shows the quantity (\ref{bufit}) for $N_2\! =\! 50k$; the fit quality for the other volumes is similar. 
The values for the fitted string susceptibility exponent decrease monotonically (within error bars) with the volume, ranging from 
$\gamma_\mathrm{str}\! =\! -0.4861\! \pm 0.0040$ for 
$N_2\! =\! 50k$ to 
$\gamma_\mathrm{str}\! =\! -0.5226\! \pm 0.0056$ for $N_2\! =\! 400k$.
They are in good agreement with the known value $\gamma_\mathrm{str}\! =\! -1/2$, especially considering the various mismatches with the minbu analysis in \cite{Ambjorn1993}. 
It shows that the bubble structure uncovered by effective homology is related to the universal properties of 2D Euclidean quantum gravity.

\section{Conclusions} 
\label{sec:summ}

We have introduced a new way of characterizing quantum geometry, as it emerges from the nonperturbative path integral in a lattice formulation in terms of dynamical triangulations. 
The key idea is to measure the Betti numbers of coarse-grained versions of the quantum geometry, using highly effective open-source tools that were developed in 
the context of TDA. We demonstrated the viability of this construction for a nontrivial example, Euclidean quantum gravity in two dimensions, where 
the nonclassical behaviour of the Betti number $\beta_2$ revealed a simple kind of ``foaminess" of its quantum geometry. We showed that this is related to 
the well-known fractal nature of the model, which was studied previously in terms of minimal-neck baby universes. This very encouraging result opens the door
for analogous investigations in higher dimensions, where we expect a potentially much richer array of effective topological features.  
The quantum observables we have introduced here provide concrete tools to get a quantitative handle on the alluring but elusive concept of quantum spacetime foam.

\vspace{0.2cm}
\subsubsection*{Acknowledgments} 

The contribution of JvdD and RL is supported by a grant in the Open Competition ENW-M Program of the Dutch Research Council (NWO), with 
grant ID https://doi.org/10.61686/IBVAP30787. The research of MS was supported by a Radboud Excellence fellowship from Radboud University in Nijmegen, Netherlands, and by an NWO Veni grant under grant ID  [\url{https://doi.org/10.61686/SUPEH07195}].
JvdD and RL thank the Perimeter Institute for hospitality.
This research was supported in part by Perimeter Institute for Theoretical Physics. Research at Perimeter Institute is supported by the 
Government of Canada through the Department of Innovation, Science and Economic Development and by the Province of Ontario through
the Ministry of Colleges and Universities. 


\end{document}